\documentclass[a4paper]{aa} 
\usepackage{graphicx}
\usepackage{natbib}
\bibpunct{ (}{)}{;}{a}{}{,}
\usepackage{txfonts}

\def\etal{et  al.\ }

\def\apj{ApJ}
\def\apjl{{ApJ\ (Lett.)}}
\def\apjs{{ApJ\ Suppl.}}

\def\aap{{A\&A}}
\def\mnras{{MNRAS}}

\def\keV {\rm keV}
\def\msol{{\rm M_{\odot}}}

\def \rv {R_{200}}
\def \mv {M_{200}}

\def\nh{{N_{\rm H}}}

\def\rs {r_{\rm s}}

\def\betamodel{$\beta$-model}
\def \etal {et al.\ }
\def \pks {\hbox{PKS 0745-191}}

\def \xmm {\hbox{\it XMM-Newton}}
\def \asca {\hbox{\it ASCA}}
\def \rosat {\hbox{\it ROSAT}}
\def \sax {\hbox{\it BeppoSAX}}
\def \chandra {\hbox{\it Chandra}}

\def \xspec {\hbox{\sc xspec}}
\def \evigweight {\hbox{\sc evigweight}}

\def \epchain{\hbox{\sc epchain}}
\def \sas {\hbox{\sc sas}}
\def \wabs {\hbox{\sc wabs}} \def \mekal {\hbox{\sc mekal}}

\begin{document}
\title{The structural and scaling properties of nearby galaxy clusters:\\
  I - The universal mass profile}

            \author{E. Pointecouteau \inst{1}, 
                    M. Arnaud \inst{1} and
                    G.W. Pratt \inst{2}}
\offprints{E. Pointecouteau, \\ \email{pointeco@discovery.saclay.cea.fr}}

         \institute{$^1$ CEA/DSM/DAPNIA Service d'Astrophysique, C.E. Saclay, L'Orme des Merisiers,
Bat. 709, F-91191 Gif sur Yvette, France\\ 
         $^2$ MPE, Giessenbachstra{\ss}e, 85748 Garching, Germany} 

\date{Received 20 Decembre 2004 ; accepted 21 January 2005} 

\abstract{We present the integrated mass profiles for a sample of ten
nearby ($z\lesssim 0.15$), relaxed galaxy clusters, covering a
temperature range of $[2-9]~\keV$,  observed with \xmm. The mass
profiles were 
derived from the observed gas density and temperature profiles under
the hypothesis of spherical symmetry and hydrostatic equilibrium. All
ten mass profiles are well described by an NFW-type profile over the
radial range from $0.01$ to $0.5 R_{200}$, where $\rv$ is the radius
corresponding to a density contrast of 200 with respect to the
critical density at the cluster redshift. A King model is inconsistent
with these data. The derived concentration parameters and total masses
are in the range $c_{200}=4-6$ and
$\mv=1.2~10^{14}-1.2~10^{15}~\msol$, respectively. Our qualitative and
quantitative study of the mass profile shape shows, for the first
time, direct and clear observational evidence for the universality of
the total mass distribution in clusters. The mass profiles scaled in units
of $\rv$ and $\mv$ nearly coincide, with a dispersion of less than
$15\%$ at $0.1~\rv$. The $c_{200}$--$\mv$ relation is consistent with
the predictions of numerical simulations for a $\Lambda$CDM cosmology,
taking into account the measurement errors and expected intrinsic
scatter.  Our results provide further strong evidence in favour of the
Cold Dark Matter cosmological scenario and show that
the dark matter collapse is well understood at least down to the
cluster scale.

}

\authorrunning{Pointecouteau \etal }
\titlerunning{The structural and scaling properties of nearby galaxy clusters - I}
\maketitle

%

\section{Introduction}
\label{sec:intro}

The Dark Matter (DM) distribution in clusters is a sensitive test of current scenarios of structure formation and of the
nature of the DM itself. Of particular interest are comparisons with
the predictions of N-body simulations of hierarchical clustering in
the currently favoured $\Lambda$CDM (Cold Dark Matter) cosmology.
Numerical simulations \citep[e.g][]{navarro97}, as well as early
semi-analytical work \citep[e.g.][]{bertschinger85}, predict a
remarkable similarity in the Cold Dark Matter density profile of
virialized halos.  Although the exact slope in the very centre is
still a matter of debate, recent high resolution simulations predict
that dark matter profiles are cusped
\citep{navarro97,moore99,diemand04,navarro04} and that the
concentration of the Dark Matter varies only slightly with system mass
\citep[e.g.][]{dolag04}.  

The strong similarity in the \rosat\ surface brightness profiles
\citep{vikhlinin99,neumann01,arnaud02}, and of the temperature
profiles of hot clusters observed with \asca\ and \sax\
\citep{markevitch98,irwin00,degrandi02} provided indirect evidence of
a universal underlying dark matter distribution. The present
generation of X--ray satellites, \xmm\ and \chandra, represent a giant
step forward in terms of resolution and sensitivity. We can now
measure precisely, through the hydrostatic equilibrium equation, the
total mass distribution in clusters.  Evidence is slowly accumulating
that CDM numerical simulations predict the correct shape of the Dark
Matter distribution, not only in massive clusters
\citep[e.g.][]{david01,allen01,arabadjis02,buote04,buote04b,allen03}, but also
in low mass clusters \citep{pratt03}.  
 This may well be true up to the virial radius, as shown up to
  $\delta=500$ by the observation of A1413 \citep{pratt02}.
The observed profiles are cusped in the centre,
and, for a few massive clusters, the inner slope has even been measured
precisely enough to distinguish between various CDM predictions
\citep{lewis03, buote04, pointecouteau04}. However, most mass studies
have been conducted on individual 'test case' clusters. Recently,
\citet{pratt05} performed the first quantitative check of the
universality of the mass profile using a sample of five clusters
observed with \xmm\ (four low mass systems compared to one massive
system).  It is necessary to extend this type of study to larger
samples, with a better temperature (i.e., mass) coverage.

In this paper, we use \xmm\ to examine the total mass profile of ten
relaxed, nearby clusters in the temperature range from 2 to 9 keV. In
a companion paper (Arnaud, Pointecouteau \& Pratt, 2005; Paper II), we
use the mass data to study the scaling properties of the mass with
temperature. In Sect.~\ref{sec:data} we present the sample, the
observations and the data processing steps. We detail the extraction
of the scientific products, from temperature and density profiles to
mass profiles.  We quantify the shape of the mass profiles in
Section~\ref{sec:mass}.  Our results are discussed and we conclude in
Sec.~\ref{sec:iss}.

We have used the currently favoured $\Lambda$CDM cosmology,
$H_0=70$~km/s/Mpc, $\Omega_m=0.3$ and $\Omega_\Lambda=0.7$, throughout
this paper.

\section{\xmm\ observations and data processing \label{sec:data}}

\subsection{The sample \label{sec:samp}}

The sample has been built to cover a wide range in temperature (i.e.
in mass) from $2$ to $9~\keV$, and constitutes 10 clusters.  We
limited the redshift range to $z\leq 0.15$, where evolution effects
are expected to be negligible.  With the exception of A478 and
PKS0745, for which there was a mosaic observation, we only considered
clusters fitting in the \xmm\ field of view,
enabling the local background to be estimated, thus limiting
systematic uncertainties on the temperature, and consequently, the mass
profiles. Since cluster size increases with temperature, this last
criterion sets a lower limit on the redshift for each temperature.
Cool and hot clusters lie in a redshift range close to this limit
($[0.04,0.06]$ and $[0.1-0.15]$ respectively).  This thus optimizes
both the cluster coverage and the statistical quality of the data. A
final selection criterion was the quality of the mass data. All
clusters in the sample have a regular X-ray morphology, indicative of
a relaxed state and allowing reliable determination of the total mass
profile through the hydrostatic equilibrium equation.

Our sample includes the sample of \citet{pratt05}: the cool clusters
A1983, A1991, MKW9, A2717, and the hot cluster A1413. We improved the
temperature coverage by adding A478 (recently studied by
\citealt{pointecouteau04}), and four clusters at intermediate and high
temperature from the \xmm\ archive which meet our criteria.  The
journal of observations is presented in Table~\ref{tab:samp}.

To minimise systematic errors in the statistical study of cluster
properties, it is important to use scientific data derived, as far as
possible, with the same procedure. Unless otherwise stated, we use the
previously published data of the cool clusters \citep{pratt03,pratt05},
A478\citep{pointecouteau04} and A1413\citep{pratt02}\footnote{We
  rescaled the published values of A1983 and A1413 to the $\Lambda$CDM
  cosmology used in this paper.}. These results were
obtained with the same general method that we use to process the four
additional clusters, although some details differ from cluster to
cluster.  The procedure is described in the next four sections.
Further details and comments on each individual target are given in
Appendix~\ref{sec:indcl}.

\begin{table}[t]
  \caption[]{Journal of observations 
    \label{tab:samp}}
  \begin{center}
    \begin{tabular}{lccccccc}
      \hline
      \hline
      Cluster & $z$  & Rev. & Mode$^{\mathrm{a}}$ & $t_{exp}$ (ksec) \\
      \hline
      A1983   & 0.0442& 400& EFF& 18/18/12\\
      A2717   & 0.0498& 558&  FF& 52/52/44 \\
      MKW9    & 0.0382& 311& EFF& 30/30/21 \\
      A1991   & 0.0586& 584&  FF& 29/29/19 \\
      A2597   & 0.0852& 179& FF&16/15/10\\
      A1068   & 0.1375& 633&  FF& 19/20/15 \\
      A1413   & 0.1430& 182& EFF& 24/25/10 \\
      A478    & 0.0881& 401,411 & EFF & 48/41/37$^{\mathrm{c}}$\\
      PKS 0745 & 0.1028& 164& -- & 10/10/--$^{\mathrm{c}}$\\
      A2204   & 0.1523& 322&  FF& 20/20/13 \\
      \hline
    \end{tabular}
  \end{center}
  Notes: $^\mathrm{a}$ EPN observation mode: FF$=$ Full Frame, EFF$=$Extended Full Frame;  $^\mathrm{b}$ Exposure time (EMOS1/EMOS2/EPN) in ksec after flare cleaning; $^\mathrm{c}$ The exposure times are given for the central pointing.
\end{table}

\subsection{Event list processing \label{evproc}}

We made use of the \xmm\
\sas\ software package, versions 5.3 or 5.4, to filter the data. Below
we detail the main data processing steps.
~\\[-1.5em]
\begin{enumerate}
\item Considering only events with $FLAG=0$ and $PATTERN \leq 12$
  (EMOS) and $PATTERN=0$ (EPN), we clean the data for soft proton
  flares  using a threshold cut method (see 
  appendix A in \citealt{pratt02}). A first screening was performed
  using light curves in 
  the high energy band ([10-12]~keV for EMOS, [12-14]~keV for EPN) using
  100s bins. After filtering using the good time intervals from this
  screening, a second screening was performed as a safety
  check-up. A second light curve built in 10s bins in a wider
  energy band ([0.3-10]~keV for EMOS, [0.3-12]~keV
  for EPN) was screened, and the event lists were filtered
  accordingly. 
\item To correct for the vignetting effect, we used the photon
  weighting method  \citep{arnaud01}. The weight coefficients were
  computed by applying   
  the \sas\ task \evigweight\ to each event file (including background
  files).  
\item The point source lists from the SOC pipeline were visually
  checked on images generated for each detector. Selected 
  point sources from all available detectors were gathered into a
  single point source list and the events in the corresponding regions were
  removed from the event lists. The area lost due to point source
  exclusion (as well as CCD gaps and bad pixels) was computed using a
  mask image. 
\end{enumerate}
~\\[-1em]

\noindent We used \xmm\ dedicated blank field datasets from either
\citet{lumb02}, or \citet{read03} to obtain a background event list
associated with each data set\footnote{Most of observations used for
the \citet{read03} blank field were centred on point sources. Their
removal, and the removal of other point sources in each field,
produced local variations of a factor of two from point to point in
the final stacked exposure map. We took this variation into account in
our analysis.}. For each cluster, the blank field event list was
recast in order to match the astrometry of the observation and was
then processed in the same way as the observation event list.  (This
included extraction of the events in the same regions as the point
sources in the observation data set.)  To account for variations in
the particle background level, we assumed that only particle signal is
collected at high energy, and used the ratio between the high energy
count rate of the observation and the blank field to renormalise the
blank field counts. The bands used were [10-12]~keV for EMOS and
[12-14]~keV for EPN. To avoid the contamination due to possible high
energy emission from the cluster, a central region ($r<5\arcmin$) was
excluded when computing the normalisation in the case of the hot
clusters.  For EPN, the out-of-time (OoT) events were considered as an
additional background component. For each observation the OoT event
list was generated using the \sas\ task \epchain\ and processed in the
same way as the observation. A normalisation factor was then used
depending on the observing mode (\emph{Full Frame} or \emph{Extended
Full Frame}).

The cleaned event list, the blank field and, in the case of the
EPN, the OoT event file of each single observation were used to
extract scientific products such as spectra and surface brightness
profiles. Background subtraction was performed using the double
subtraction process fully described in \citet[][ Appendix]{arnaud02b},
and involves subtraction of the normalised blank field data, and
subsequent subtraction of the Cosmic X-ray background residual
estimated from a region free of cluster emission.

\subsection{The density profile \label{sec:neprof}}

For each cluster, an azimuthally-averaged, background subtracted
surface brightness (SB) profile was computed in the soft energy band
([0.3-2.]~keV in the present work) for each available
detector\footnote{We also computed background subtracted SB profiles
in the high energy band (we recall: [10-12]~keV for EMOS and
[12-14]~keV for EPN) to check that there were no significant
residuals, i.e., that the normalisation of the blank field background
was correct, and that it was indeed constant with radius.}.  The
profiles of all detectors were then summed together into a total SB
profile and rebinned with logarithmic radial binning and a minimum S/N
ratio of $3\sigma$.  The profile was corrected for radial variations
of the emissivity (e.g due to abundance or temperature gradients) in
the energy band considered \citep[see][ for details]{pratt03}.  This
corrected profile is thus proportional to the emission measure along
the line of sight.

The final SB profiles were fitted using parametric analytic models of
the gas density profile, converted to an emission measure profile and
convolved with the PSF spatial response
\citep{ghizzardi01,ghizzardi02}.  We considered various parametric
forms and empirically chose the model best fitting the data using the
$\chi^2$ statistic as a measure of the goodness of fit.  The models
included a double \betamodel\ (the BB model defined in
\citealt{pratt02}),
a modified double \betamodel, which allows a more concentrated gas
density distribution towards the centre (the KBB model used by
\citealt{pratt02} for A1413), and the sum of three $\beta$-models in
which a common value of $\beta$ is assumed to ensure smooth behaviour
at large radii (the BBB model used by \citealt{pointecouteau04} to
model the SB profile of A478). For the clusters newly analysed here,
the best fitting model was either a KBB model (A2597, A1068 and \pks)
or a BBB model (A2204).
Over the whole sample the reduced $\chi^2$ values vary from 1.1 (for
A478), to 1.5 (for A1068), and thus even the best parametric model
leads in some cases to a formally unacceptable fit. This is linked to
the very small statistical errors on each measurement. The actual
discrepancies between the model and the data remain small: adding a
few percent (2 to 5\%) of systematic error while fitting the data
always leads to acceptable $\chi^2$ values.

\subsection{The temperature profile}\label{sec:ktprof}

\begin{figure*}[t]
\begin{center}
\includegraphics[width=\columnwidth]{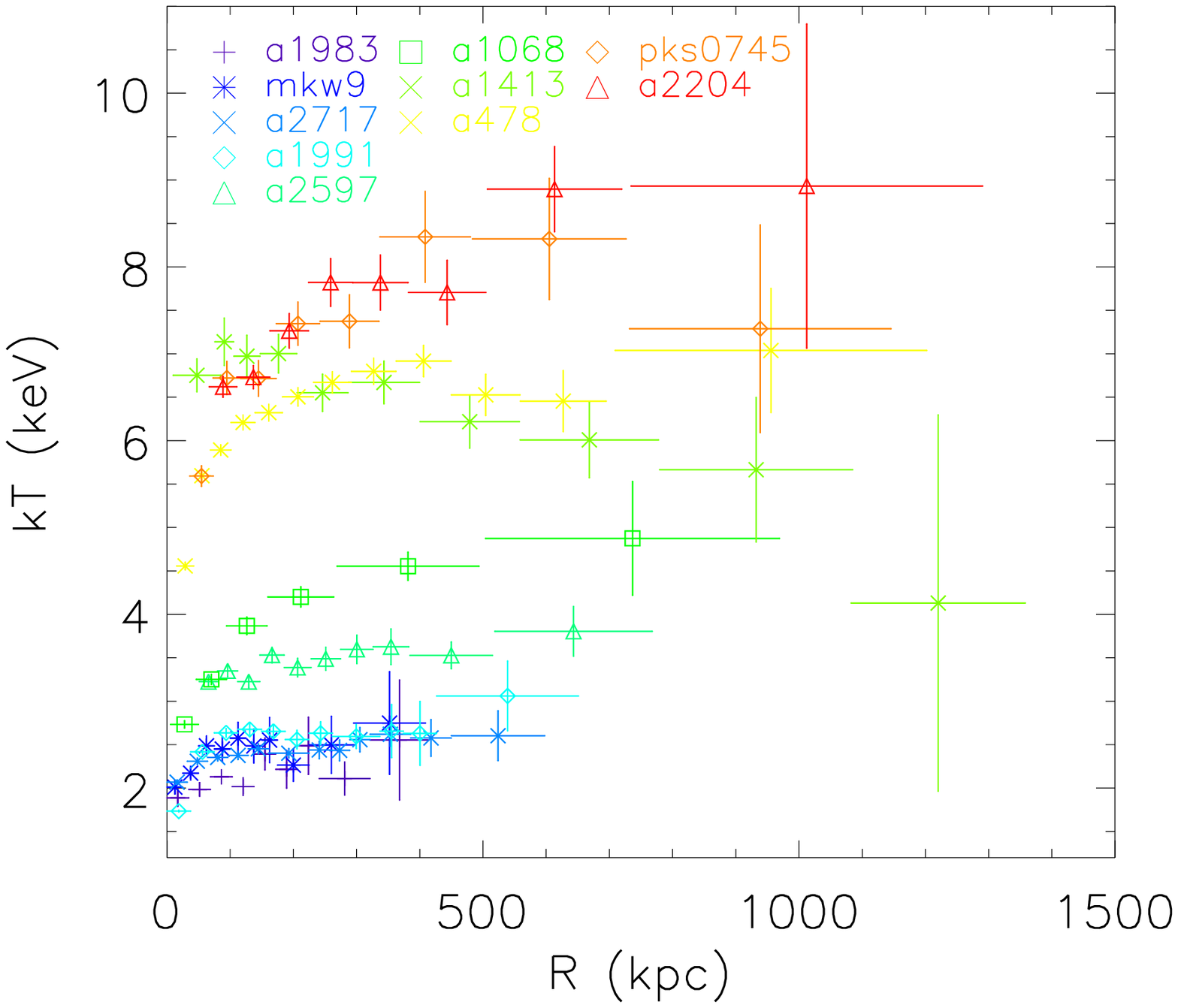}
\includegraphics[width=\columnwidth]{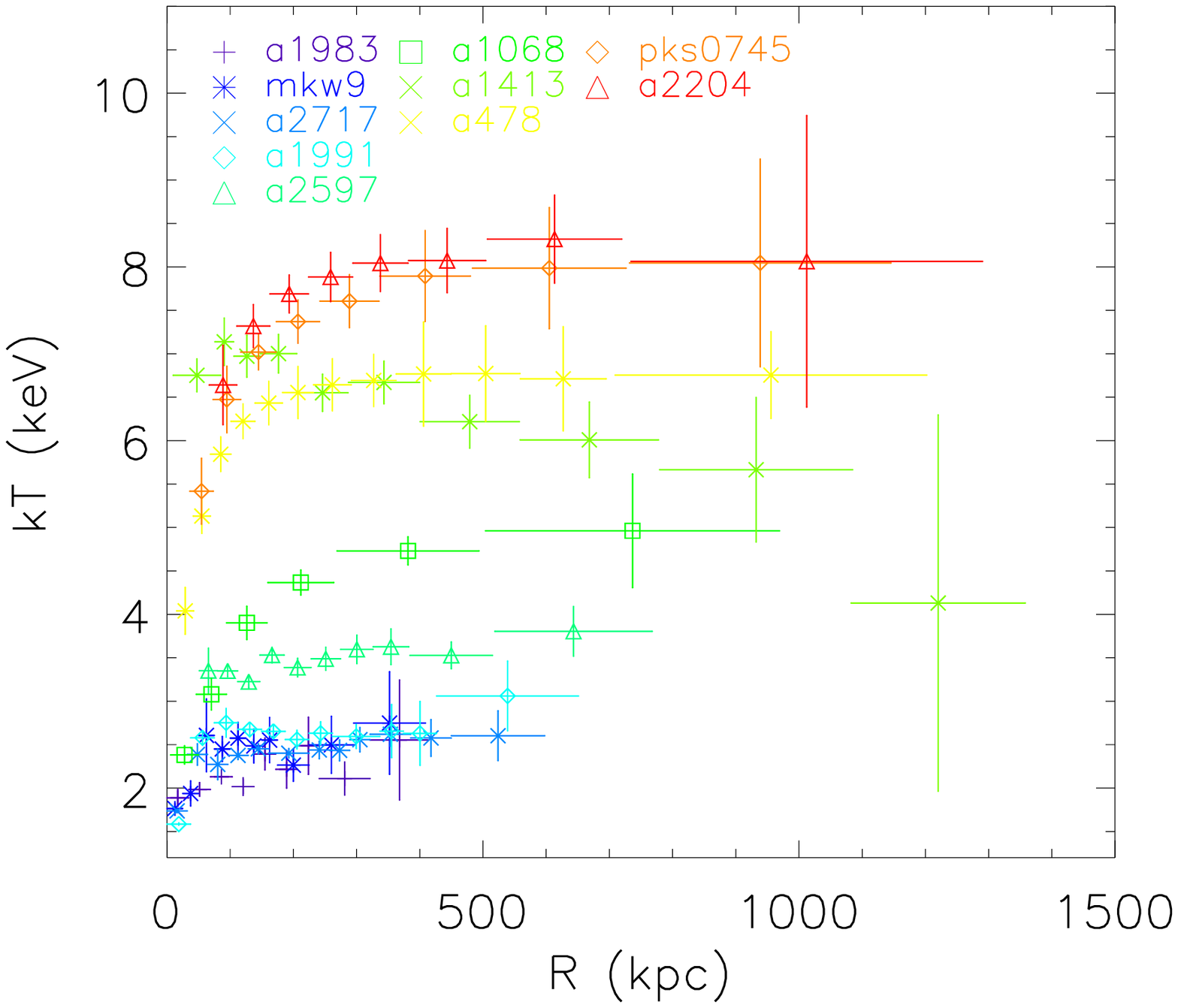}
\caption{{\footnotesize Left: Annular (projected) cluster temperature
    profiles. Right: Deprojected, PSF-corrected temperature profiles.}}\label{fig:ktprof} 
\end{center}
\end{figure*}

Concentric annular regions were defined from the total background
subtracted SB profile of the cluster 
the clusters newly analysed here, we used the following empirical
method to define the annuli from which the spectra were extracted.
We started with a minimum bin size of $15\arcsec$ (e.g about the
  HEW of the XMM-Newton PSF) and then increase the bin size by a
  logarithmic factor of 1.05. This is sufficient in the central parts
  to keep a good S/N, but not in the outer regions. We thus further
  impose, in the outer regions, that the numbers of cluster counts per
  bin is approximately constant, within $5\%$.
 
For each annular region, a background subtracted spectrum was
extracted for each available detector. The spectra from each detector
were then simultaneously fitted using \xspec\ \citep{karnaud96} with
an absorbed, redshifted single temperature plasma model
(\wabs*\mekal). Except in the case of A478 \citep[see][ for
details]{pointecouteau04}, after checking that the $\nh$ value agreed
with the galactic 21 cm value (from \citet{hartmann99}), this
parameter was frozen. Thus we derived an annular temperature profile
for each cluster in the sample, as shown in the left panel of
Fig.~\ref{fig:ktprof}. 
All profiles (except that of A1413) have the same generic shape: a
negative gradient towards the centre and a roughly flat external
plateau. A1413 is the only cluster for which we observe a
significantly decreasing temperature at high radii (of about 20\%).

To derive the true radial temperature profile (needed to compute the
mass profile) we should take into account projection and PSF
effects. For the typical temperature profile shape we have obtained,
these effects depend most strongly on the magnitude of the gradient in
the centre.  In addition, a strong temperature gradient in the central
regions is usually associated with a strongly peaked surface
brightness profile, further increasing the PSF blurring\footnote{The
  effects of the PSF correction are most noticeable for A2204
and A478.}. Reconstructing the radial
temperature profile in these cases is not a trivial task. General
non-parametric methods, such as simultaneous fitting of annular
spectra, amplify the noise considerably.  This yields radial
temperature profiles with unphysically large fluctuations,
particularly when both PSF and projection effects are important
\citep[see discussion in][]{pointecouteau04}.

The clusters newly analysed in the present work have quite strong
`cooling flows'. We thus applied the method developed for A478 by
\citet{pointecouteau04}, in which the radial temperature profile is
derived from the annular temperature profile in the following manner.
The noise amplification problem is avoided by using smooth parametric
representation of the annular temperature profile. We used the
function given by \citet{allen01}: $T(r) = T_0+T_1\,
\left[\,\frac{(r/r_c)^\eta}{1+(r/r_c)^\eta}\,\right]$ to fit the
annular profile. The best fitting model profile is then corrected for
both the projection and PSF effects, assuming that the annular
temperatures are emission weighted temperatures \citep[see][ for
details]{pointecouteau04}. To estimate the errors we repeated the
procedure 1000 times, using a Monte Carlo method that randomizes the
annular profile based on the observed errors. Because using a specific
functional form effectively limits the allowed profiles, the 
standard deviation of the corrected temperature at a given point is
occasionally smaller than the error on the annular temperature. When
this was the case, we kept the observed error.

For the poor systems, which have a more modest central temperature
gradient, the radial temperature profile was estimated as described in
\citet{pratt03,pratt05}. For A1413 PSF and projection effects proved
to be negligible \citep{pratt02}. The deprojected, PSF corrected 
profiles of all the clusters are shown in the right panel of
Fig.~\ref{fig:ktprof}. 
A detailed discussion of the shape of these temperature profiles is
  beyond the scope of this paper.

\subsection{The total mass profile \label{sec:mprof}}

\begin{figure*}[!t]
\centering
\hspace*{-0.5em}
\includegraphics[width=\columnwidth]{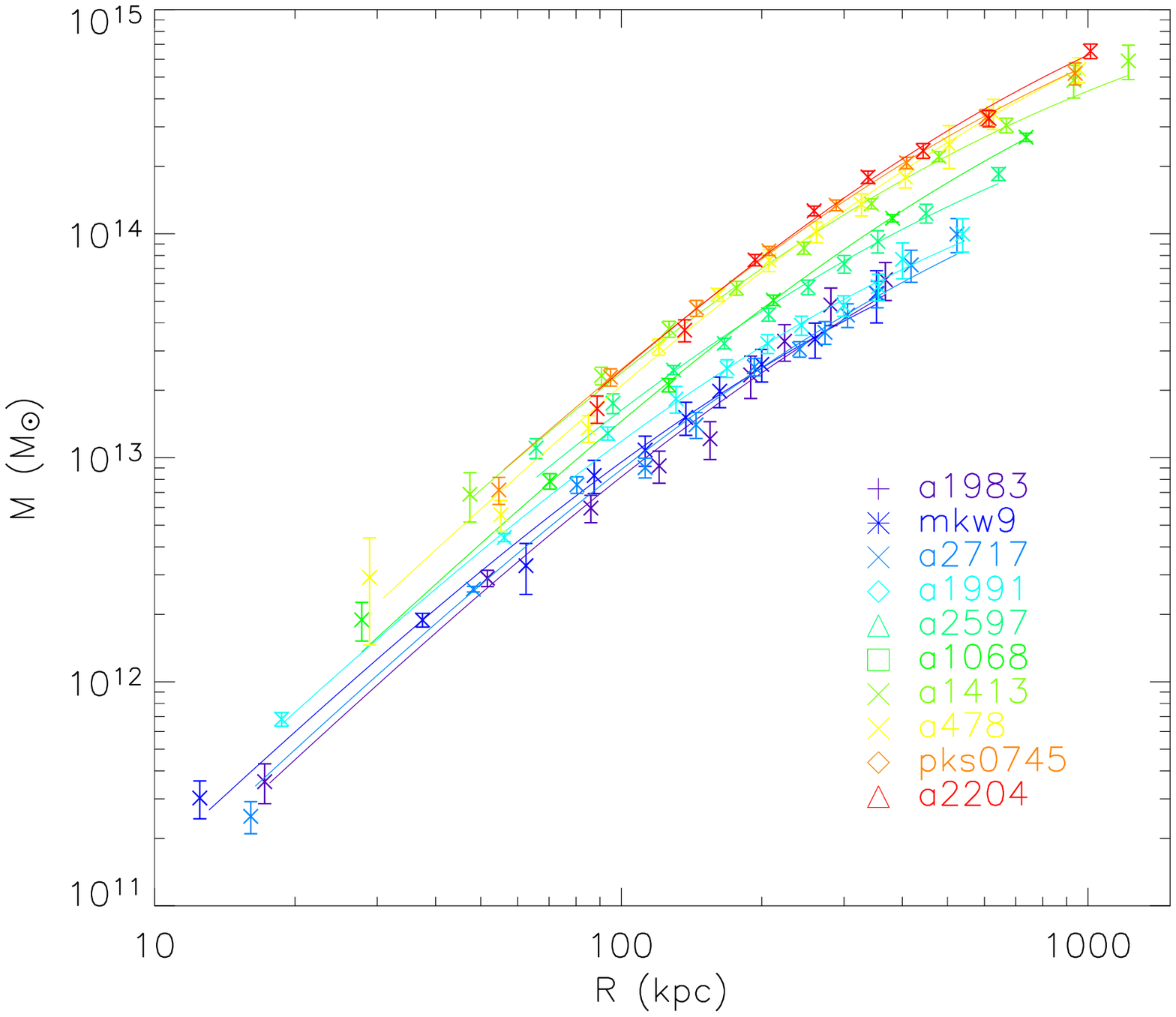}
\hspace*{1em}
\includegraphics[width=\columnwidth]{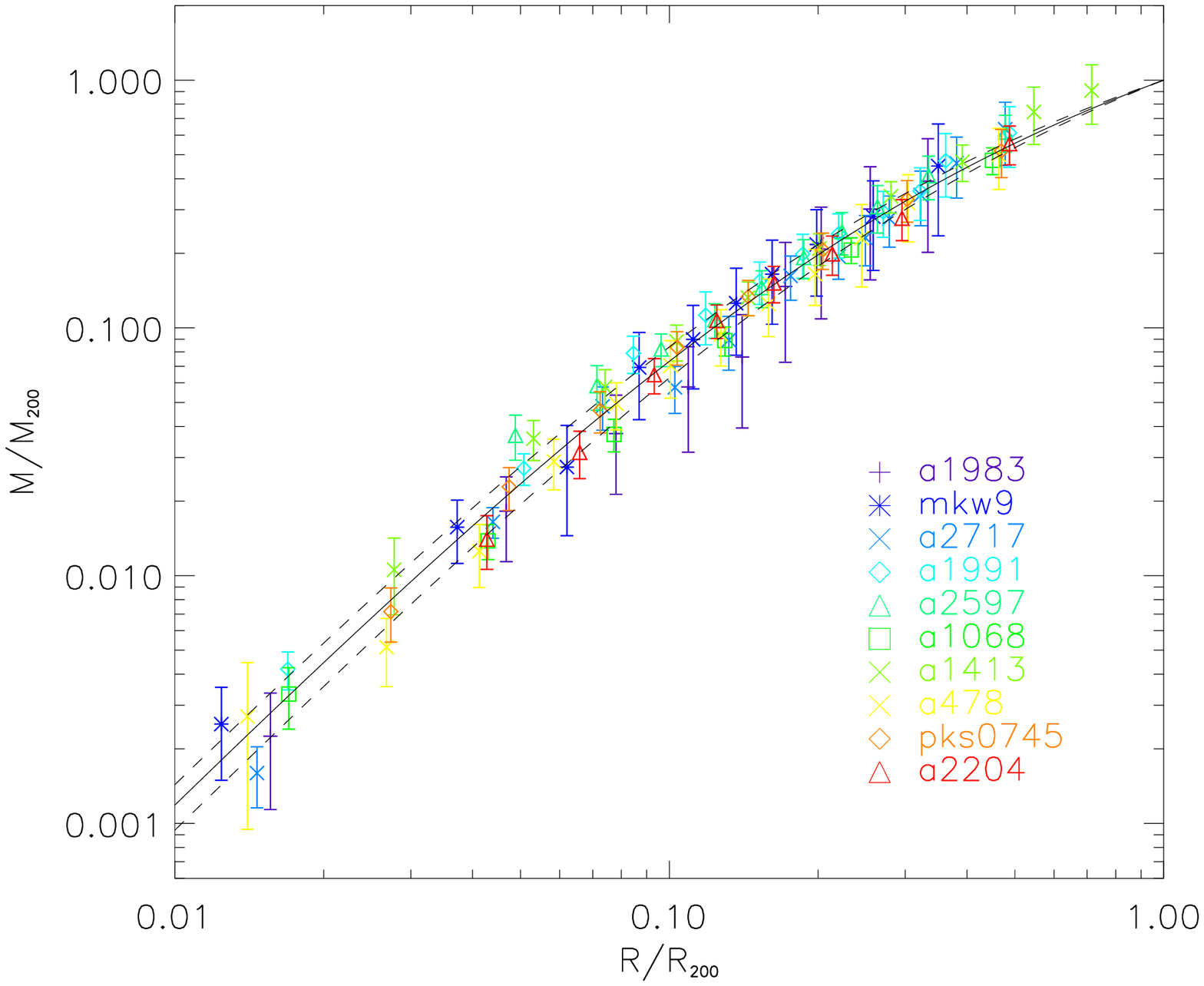}
\caption{{\footnotesize Left panel: Integrated total mass profiles
plotted in units of physical radius (kpc). The solid lines are the
best fitting NFW profiles as detailed in Table~\ref{tab:mdel}.  Right
panel: Scaled mass profiles of all clusters. The mass is scaled to
$\mv$, and the radius to $\rv$, both values being derived from the
best fitting NFW model. The solid black line corresponds to the mean
scaled NFW profile and the two dashed lines are the associated standard
deviation.}}\label{fig:mprof}
\end{figure*}

The mass profile for each cluster was derived from the best fitting
density profile and the deprojected, PSF corrected temperature profile
under the assumptions of hydrostatic equilibrium and spherical symmetry:
\begin{equation}
M(r) = - \frac{kT\ r}{{\rm G} \mu m_p}   \left[ \frac{d \ln{n_{\rm
g}}}{d \ln{r}} + \frac{d \ln{T}}{d \ln{r}} \right].
\label{eq:he}
\end{equation}

\noindent The mass, and associated errors, were then computed at each
radius of the temperature profile\footnote{We use in fact the weighted
effective radius of each annulus \citep{lewis03}.} using a Monte Carlo
method that randomises the temperature profile based on the observed
errors \citep{pratt03}. A cubic spline interpolation is used to
compute the temperature derivative at each radius. Only mass profiles
which monotonically increase with radius were kept (1000 in
total). The errors due to the errors on the density gradients, which
are negligible with respect to those from the temperature, were
then added. The mass profile of A1413 was originally derived using a
slightly different Monte Carlo method \citep{pratt02}. For homogeneity
of the present study, we re-derived the mass profile of this cluster
with the present method, allowing us to extend the profile to lower
radii\footnote{\citet{pratt02} used the original Monte Carlo method of
\citet{neumann95}. This method uses a `diffusive' process to calculate
random temperatures at equally spaced radii. It has two parameters,
the radial step and a window parameter, which are optimised to ensure
smooth random temperature profiles. In this original version of the
code, the radial step fixed the minimum radius of the mass profile. In
the present method it is simply the radius of the first temperature
annulus.}. The new mass profile is perfectly consistent with the
published one.

Each cluster has been checked for the presence of structure such as
cold fronts, hot bubbles, ghost cavities, or other effects which could
affect the mass determination. Details of each cluster are given in
Appendix~\ref{sec:indcl}.

\section{The shape of the mass profile \label{sec:mass}}

\begin{table*}
  \caption[]{{\footnotesize Results for the NFW mass profile fits.} }
\begin{center}
  \label{tab:mdel}
 \begin{tabular}{lccccccl}
  \hline
 Cluster &$c_{200}$& $\rv $ (kpc)& $\mv$ ($10^{14}\, \msol$)& $R_{500} $ (kpc)& $M_{500}$ ($10^{14}\, \msol$) & $\chi^2_{NFW}$(dof) &[$\chi^2_{king}$, $\chi^2_{MQGSL}$]$^\textrm{(a)}$ \\
 \hline
     A1983  &  $3.83\pm0.71$  & $1100\pm 140$  & $1.59\pm 0.61$  & $717\pm 79$ &$1.09\pm0.37$   &9.1(7) &[31, 12] \\
      MKW9  & $5.41\pm0.67$   & $1006\pm  84$  &  $1.20\pm 0.30$ & $668\pm 51$ &$0.88\pm0.20$  &3.6(8) &[24, 5.2]\\
     A2717  & $4.21\pm0.25$   & $1096\pm  44$  & $1.57\pm 0.19$  & $717\pm 26$ &$1.10\pm0.12$   &16(10) &[48, 52]\\
     A1991  & $5.78\pm0.35$   & $1106\pm  41$  & $1.63\pm 0.18$  & $737\pm 25$ &$1.20\pm0.12$   &10(9) &[82, 33]\\
     A2597  & $5.86\pm0.50$   & $1344\pm  49$  &  $3.00\pm 0.33$ & $897\pm 29$ &$2.22\pm0.22$  &14.6(8) &[55, 6.8]\\
     A1068  & $3.69\pm0.26$   & $1635\pm  47$  &  $5.68\pm 0.49$ & $1060\pm 26$ &$3.87\pm0.28$ &2.5(4) &[62, 9.8] \\
    A1413  & $5.82\pm0.50$    & $1707\pm  57$  &  $6.50\pm 0.65$ & $1129\pm 33$ &$4.82\pm0.42$  & 8.0(8) &[45, 3.7]\\
      A478  & $4.22\pm0.39$   & $2060\pm 110$  & $10.8\pm 1.8$   & $1348\pm 64$ &$7.57\pm1.11$ &9.5(10) &[16, 30]\\
   PKS0745 & $5.12\pm0.40$    & $1999\pm  77$  &  $10.0\pm 1.2$  & $1323\pm 45$ &$7.27\pm0.75$  &2.3(6) &[20, 20]\\
     A2204  &  $4.59\pm0.37$  & $2075\pm  77$  & $11.8\pm 1.3$   & $1365\pm 44$ &$8.39\pm0.81$  &9.7(6) &[14, 21]\\
 \hline
 \end{tabular}
\end{center}
$\textrm{(a)}$ Chi-square obtained for the best fit of a king and a MQGSL model respectively (see Sect.~\ref{sec:mmod})
\end{table*}

\subsection{Mass profile modelling \label{sec:mmod}}

For each cluster, three mass models were fitted to the data: (i) a
King isothermal sphere profile; (ii) a standard NFW profile
\citep{navarro97}; and (iii) an MQGSL profile \citep{moore99}. Our data
indicate that an isothermal sphere model (i.e., a profile with a core)
is not a good representation of the mass distribution in these
clusters. Dropping too rapidly in the centre and flattening in the
outer regions, the reduced $\chi^2$ obtained from King model fits
ranged from 1.65 for A478 to 15.6 for A1068. It is rejected with a
minimum 91\% confidence level (A478). In contrast, the reduced
$\chi^2$ obtained from NFW model fits varied from from 0.4 (\pks) to
1.8 (A2597), while the MQGSL profile yielded reduced $\chi^2$ of 0.5
(A1413) to 5.21 (A2717). We note that the chi-squared value is very
sensitive to the central points. In some cases, the mass errors on
these points may be underestimated due to the procedure used for PSF
and projection effects correction, or to systematic errors we are not
able to quantify and are therefore unable to take into account.

In all cases barring A1413 \citep{pratt02} and A2597, the NFW profile
proved to be a better fit than the MQGSL model, and \emph{de facto} to
be the best representation of our current data. In the cases of A1413
and A2597, the improvement in $\chi^2$ when an MQGSL profile is used
is very small. Thus, to keep our approach coherent, we decided to use
the NFW fit as a parametric representation of the mass profile of each
cluster.

The NFW model, where the density is $\rho(r) \propto [(r/\rs)
(1+r/\rs)]^{-1}$, has two free parameters: the scaling radius $\rs$,
and a normalisation parameter. The model can be equivalently expressed
in terms of the concentration parameter $c_{200} = r_{200}/\rs$ and
the total mass $\mv$. $\mv$ is the mass corresponding to a density
contrast of $\delta=200$, i.e. the mass contained in a sphere of
radius $\rv$, which encompasses a mean density of 200 times the
critical density at the cluster redshift: $\rho_{\rm c}(z)= 3
E(z)^2{\rm H_0}^2 /8\pi{\rm G}$, where $E^{2}(z)=\Omega_{\rm m}(1+z)^{3}
+\Omega_{\Lambda}$.  In numerical simulations, this sphere is found to
correspond roughly to the virialised part of clusters.  The results of
the best NFW fits are detailed in Table~\ref{tab:mdel}, and the best
fitting profiles are shown in Fig.~\ref{fig:mprof}.

\subsection{Scaled mass profiles}

The left panel of Fig.~\ref{fig:mprof} shows the mass profiles (and
best-fitting NFW models) plotted in physical units (kpc). Not
surprisingly, there is a continuous increase in mass with cluster
temperature, reflecting the temperature coverage of the sample. These
unscaled mass profiles already show signs of an underlying similarity
in the matter distribution. The right panel of Fig.~\ref{fig:mprof}
shows the scaled mass profiles, where we express the radius in terms
of $\rv$ and the mass in terms of $\mv$, these values being derived
from the best fitting NFW model of each cluster. The scaled mass
profiles cover a wide range of radii, from about $0.01~\rv$ to
$0.7~\rv$, and are particularly well constrained between $0.1~\rv$ and
$0.5~\rv$.  The agreement between the scaled profiles is remarkable,
reflecting the similarity in shape of the profiles.  The average of
all best fitting NFW models is shown as a black line, with dashed lines
representing the mean plus or minus the standard deviation.  The
dispersion is small and virtually identical to that derived by
\citet{pratt05} from their smaller sample: we obtained a dispersion of
$7.4\%$ at $0.3\rv$ and $14.3\%$ at $0.1\rv$ (compared to the $8\%$ and
$15\%$ found by those authors).

\subsection{Variation of the concentration parameter with mass}

\begin{figure}[t]
\begin{center}
\includegraphics[width=\columnwidth]{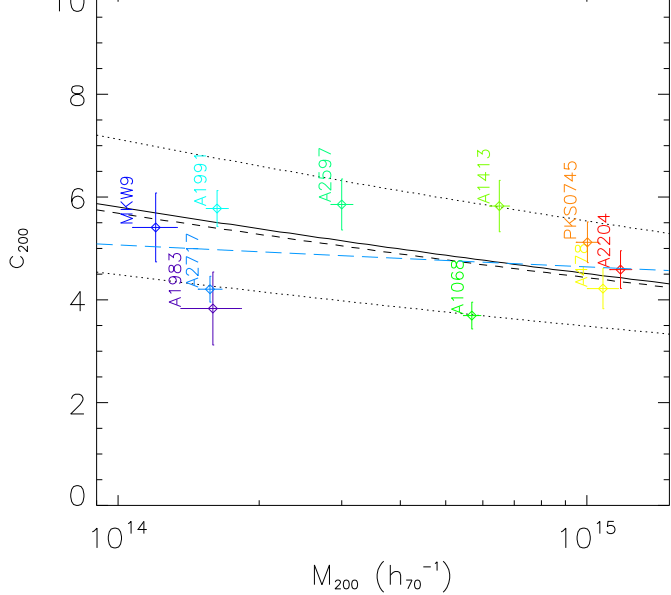}
\caption{{\footnotesize Concentration parameter $c_{200}$ versus the
    cluster mass $M_{200}$. The solid line represents the variation of
    $c_{200}$ for clusters at $z=0$ from the numerical simulations of
    \citet{dolag04}. The dotted lines are the standard deviation
    associated with this relation. The dashed line represents the same
    relation at a redshift of $z=0.15$( the maximum redshift for our
    sample). The long-dashed line stands for our best fit of the
    $c_{200}-M_{200}$ (see text).}}\label{fig:cm}
\end{center}
\end{figure}

Structure formation models do not in fact predict a strictly universal
matter distribution in clusters. A weak variation in the concentration
is expected from low to high mass clusters, reflecting differences in the
formation epochs of low and high mass haloes
\citep{navarro97,bullock01,dolag04}.  Building on the work of
\citet[][ their Fig.~12]{pratt05}, we can investigate the relation between
the concentration parameter and the cluster mass using our extended
sample. Figure~\ref{fig:cm} presents the $c_{200}-M_{200}$ relation
found by \citet{dolag04} $(1+z)\bar{c}=c_0(M/M_0)^\alpha$ with
$\alpha\sim -0.1$ in a $\Lambda$CDM cosmology with $\sigma_8=0.9$. All
ten points agree with the predicted relation, taking into account its
intrinsic dispersion and the measurement errors. 

We then performed a linear regression fit in the
$\log{c_{200}}$--$\log{M_{200}}$ plane, taking into account the
uncertainties on each quantity. The resulting slope of $\alpha = -0.04
\pm 0.03$ ($\chi^2$ (dof) = 42(8)) is poorly constrained, due in part
to the limited size of the sample and the large relative uncertainties
in the concentration parameter.  The result is however compatible with
the intrinsic dispersion of theoretical predictions. The best fitting
$c(M)$ relation is shown as the long dashed line in
Fig.~\ref{fig:cm}.

\section{Discussion and conclusion  \label{sec:iss}}

We have measured the total mass profile of ten clusters from $0.01\rv$
up to $0.5 R_{200}$ using \xmm\ observations. Our sample has an
excellent temperature coverage and covers an order of magnitude in
mass from $\mv = 1.2~10^{14}~\msol$ to $1.2~10^{15}~\msol$. Our study
confirms previous \xmm\ and \chandra\ studies conducted on individual
targets (see Sec. ~\ref{sec:intro}), and extends the initial
statistical study of cluster mass profile structure by
\citet{pratt05}.

We have found that the NFW profile is a good representation of the ten
observed mass profiles, and that in all cases the isothermal sphere
model (i.e a profile with a core) is rejected at high confidence. In
other words, we confirm the cusped nature of the Dark Matter profile,
as predicted by CDM simulations of hierarchical structure formation,
over the temperature/mass range of the present sample. The mass
profile shape is close to universal, again as predicted, with a
dispersion of less than $15\%$ at $0.1\rv$ in the scaled mass
profiles. The shape is quantitatively consistent with theoretical
predictions. The variation of the observed concentration parameters
with mass is in line with the predictions, taking into account the
measurement errors and the expected intrinsic scatter. However, our
sample is still too small to draw any firm conclusions on the exact
form of the $c(M)$ relation.  Taken together, our results provide
further strong evidence in favour of the Cold Dark Matter cosmological
scenario, and show that the physics of the Dark Matter collapse is
well understood.

The exact inner slope of the CDM distribution in haloes remains an
important theoretical issue \citep{diemand04,navarro04}. Few of our
mass profiles have the required radial coverage and statistical
quality in the central parts to allow us to firmly distinguish between
an NFW-type profile and other types of cusped DM profile (e.g., the
mass profile of A478). We caution also that the very central parts of
clusters are regions of complex phenomena (hot bubbles, ghost
cavities, cold fronts, interaction with the central galaxy, etc) which
are still not well understood.  Their effect on the ICM may challenge
the hypothesis of hydrostatic equilibrium, and therefore the
reliability of the X-ray mass estimate.

Observations are also needed of the outskirts of clusters. To date,
those regions are basically unknown to observers, and can only be
investigated with numerical simulations. Study of these regions is
clearly needed to advance our understanding of structure formation and
evolution.


\begin{acknowledgements}
The authors thank the anonymous referee for his remarks and comments.
This research has made use of the \xmm\ archives and of the SIMBAD
database, operated at CDS, Strasbourg, France.
EP acknowledges the financial support of CNES (the French space
agency). GWP acknowledges funding from a Marie Curie Intra-European
Fellowship under the FP6 programme (Contract
No. MEIF-CT-2003-500915). The authors thank Doris Neumann for
interesting discussions and Nabila Aghanim for her help in the early
analysis of A2204.
\end{acknowledgements}

\appendix
\section{Cluster specifics \label{sec:indcl}}
For each cluster in our sample, we searched the literature for special
features such as ICM bubbles, ghost cavities, cold fronts or other
phenomena which could disturb the relaxed structure of the
intra-cluster medium and therefore invalidate out hypothesis of
spherical symmetry and hydrostatic equilibrium.  We particularly
investigated existing Chandra observations of central substructure in
clusters, capitalising on its high spatial resolution. We used this
information to check our mass profiles and if needed to exclude one
(or more) point(s) containing such substructure.  Details and notes
are now given for individual clusters.
\\[-1.5em]
\begin{description}
\item[A478] -- %
  From the Chandra observations, \citet{sun03} reported the presence
  of X-ray cavities within the 15 central kpc. The recent \xmm\
  observation of A478 \citep{pointecouteau04} excluded
  this inner area for the mass profile computation. We used the
  published mass profile of \citet{pointecouteau04}.
\item[A1068] -- %
  We chose this cluster for its intermediate temperature of about
  4-5~keV. The data for A1068 were retrieved from the \xmm\ archive.
  According to the Chandra observation detailed in \citet{wise04}, no
  special inner structure has been seen.
\item[A1413] -- %
  This cluster has been studied in detail by \citet{pratt02}. Among
  the sample it is the only hot cluster without a cool core,  and the
  only cluster with a firmly
  observed decrease in temperature at large radii. It is
  also the cluster observed up to the largest radius ($0.7 R_{200}$). The
  existing (public) Chandra observation of this cluster has a too low
  an exposure time to show clearly whether there is structure in the
  cluster centre. 
\item[A1983] -- %
  \citet{pratt03} published an extensive study of the \xmm\
  observations for this cluster.  We used all the mass points published
  by those authors. No Chandra X-ray observation
  is yet reported for this cluster.
\item[A1991] -- %
  \citet{mcnamara04}, using Chandra, have reported low energy knots
  ($kT\sim 0.8$~keV) located at $\sim 10$~kpc from the cluster
  centre. The first bin of the temperature (and thus mass) profile, used
  by \citet{pratt05} in their \xmm\ study of A1991, extends up to
  38~arcsec. The knot structure seen by Chandra is included in this
  bin and is therefore diluted within the cluster emission. Keeping or
  removing the first point in our mass analysis leads to mass
  modelling results which are compatible at a 1$\sigma$ level. The NFW
  fit being slightly better when 
  using all the points, we choose to use the whole mass profile
  published by \citet{pratt05}.
\item[A2204] -- %
  This is a very well relaxed cluster at a redshift of $z=0.1523$
  showing a very peaked emission profile toward the centre. The
  cluster hosts a very strong radio source at its centre, and recently
  \citet{sanders04} have pointed out complex structure in the very
  central part of the cluster using the Chandra satellite. In
  addition, as
  discovered in many other clusters, two cold fronts were observed at
  radii of about 30 and 80~kpc.  While we computed the mass profile
  down to small radii for this cluster, we excluded in the mass
  analysis the two inner annuli, the outer radius of which extend up to
  109~kpc, in order to avoid invalidating the assumption of
  hydrostatic equilibrium. 
\item[A2717] -- %
  Another poor cluster from \citet{pratt05}. No high resolution X-ray
  imaging has been performed to date on this cluster. However, a work
  based on ROSAT/PSPC and APM optical data aiming to find substructure
  in a sample of clusters \citep{kolokotronis01}, reported
  substructure in the optical for A2717. It is undetectable in
  this X-ray observation. No significant change occurring when
  excluding the central 
  point from the fit, we used the entire computed mass profile.
\item[A2597] -- %
  This cluster was taken from the \xmm\ archive, and was chosen
  because of its intermediate temperature of about 3-4~keV. A2597 has
  been studied with the Chandra satellite, exhibiting ghost cavities in
  its centre at a radius of about 30~kpc \citep{mcnamara01}. Moreover,
  the Chandra SB profile and temperature profile seems to be quite
  disturbed up to a few tens of kpc. We therefore choose to ignore the
  two first data points of our mass profile and to only keep points
  with inner radii larger than 50\arcsec.
\item[MKW9] -- %
  This cluster is the least regular of the three cool systems recently
  studied by \citet{pratt05}. However, lacking any external
  information on its central structure we use their published mass
  profile. 
\item[\pks] -- %
  A previous analysis of \xmm\ data was published by
  \citet{chen03}. However, we decided to reprocess the whole dataset
  to keep the data processing in our
  sample coherent.  The data for this cluster are strongly disturbed by
  important flaring periods. After flare cleaning, we kept only the MOS
  data, the EPN data being still very noisy. Our temperature profile
  is in total agreement
  with that of \citet{chen03}, and also with the measurement by BeppoSAX at
  large radii \citep{degrandi99}. Note that the overall normalisation
  of our temperature profile is
  somehow lower to that obtained by \citet{hicks02} with the Chandra
  satellite.  Indeed, these authors obtained a plateau temperature of
  about $10.5$~keV above $200$~kpc from the centre against our value
  of $\sim8$~keV. Regarding the SB profile from the Chandra
  observation, which exhibits a sharp central cusp (mainly due to the
  cD galaxy), we improved the fit of the mass profile by excluding our
  central bin which extends up to 34\arcsec.
\end{description}


\end{document}